\documentclass[12pt,epsf]{article}
\usepackage[pdftex]{graphicx}
\pdfoutput=1
\usepackage{amsmath,amssymb}
\usepackage{bm}
\usepackage{amsfonts}
\usepackage{mathrsfs}
\usepackage{ascmac}
\usepackage{braket}
\graphicspath{{./fig/}}
\usepackage{comment}
\usepackage{cite}

\newcommand{\beq}{\begin{equation}}
\newcommand{\eeq}{\end{equation}}
\newcommand{\bea}{\begin{eqnarray}}
\newcommand{\eea}{\end{eqnarray}}

\newcommand{\ba}{\begin{aligned}}
\newcommand{\ea}{\end{aligned}}

\def\pe2{p_E^2}

\begin{document}
\setlength{\baselineskip}{0.7cm}
\begin{titlepage} 
\begin{flushright}
NITEP 138
\end{flushright}
\vspace*{10mm}%
\begin{center}{\LARGE\bf
Cosmological Collider Signals of \\
\vspace*{2mm}
Non-Gaussianity from Higgs boson in GUT
}
\end{center}
\vspace*{10mm}
\begin{center}
{\Large Nobuhito Maru}$^{a,b}$ and 
{\Large Akira Okawa}$^{a}$ 
\end{center}
\vspace*{0.2cm}
\begin{center}
${}^{a}${\it 
Department of Physics, Osaka Metropolitan University, \\ 
Osaka 558-8585, Japan}
\\
${}^{b}${\it Nambu Yoichiro Institute of Theoretical and Experimental Physics (NITEP), \\
Osaka Metropolitan University, 
Osaka 558-8585, Japan} 
\end{center}
\vspace*{1cm}

\begin{abstract} 
Cosmological Collider Physics gives us the opportunity to probe high-energy physics 
 from observing the spacetime fluctuations generated during inflation imprinted on the cosmic microwave background. 
In other words, it is a method to investigate physics on energy scales that cannot be reached by terrestrial accelerators 
 by means of precise observations of the universe. 
In this paper, we focus on the case where the GUT scale is close to the energy scale of inflation, 
 and calculate three point function of inflaton by exchanging the Higgs boson in GUT at tree level. 
The results are found to be consistent with the current observed restrictions on non-Gaussianity 
 without a drastic fine tuning of parameters, 
and it might be possible to detect the signature of the Higgs boson in GUT 
 by 21cm spectrum, future LSS and future CMB depending on our model parameters. 
 \end{abstract}
\end{titlepage}

\section{Introduction} 
The Standard Model of elementary particles can explain a wide variety of elementary particle phenomena and has acquired reliability. 
On the other hand, there are certainly some phenomena that cannot be explained by the Standard Model, 
for example, the mechanism of generating tiny neutrino masses, the origin of dark matter and dark energy, 
and the hierarchy problem. 
To solve these problems, various theories beyond the Standard Model have been considered, 
 such as extra dimensional scenario, supersymmetry, and Grand Unified Theory (GUT). 
GUT is the theory that unifies three of the four fundamental interactions that exist in nature: 
 strong interaction, electromagnetic interaction, and weak interaction. 
The renormalization group method suggests that these three interactions are unified at a certain energy scale. 
Since the Standard Model is described by the Weinberg-Salam model, which unifies the electromagnetic and weak interactions, 
 it is natural that the strong interaction is also unified. 
However, the energy scale of the GUT is expected to be about $10^{15}$ GeV, 
 and it is difficult to verify such a high energy theory by terrestrial accelerators. 
For this reason, Cosmological Collider Physics has attracted interest 
\cite{1, 2, 3, 4, 5, 6, 7, 8, 9, 10, 11, 12, 13, 14, 15, 16, 17, 18, 19, 20, 
21, 22, 23, 24, 25, 26, 27, 28, 29, 30, 31, 32, 33, 34, 35, 36, 37, 38, 39, 
40, 41, 42, 43, 44, 45, 46, 47, 48, 49, 50, 51, 52, 53, 54, 55, 56, 57, 58, 59, 
60, 61, 62, 63, 64, 65, 66, 67, 68, 69, 70, 71, 72, 73, 74, 75, 76, 77, 78, 79, 
80, 81, 82, 83, 84, 85, 86, 87, 88, 89, 90, 91, 92, 93, 94}. 
Cosmological Collider Physics is a field that obtains information on high energy elementary particles 
 by observing quantum fluctuations in space-time stretched by inflation through the cosmic microwave background radiation. 
That is, precise observation of the universe can provide information on elementary particles 
 in high energy which cannot be reached by terrestrial accelerators.

\begin{figure}[htbp]
 \begin{center}
  \includegraphics[width=10cm,]{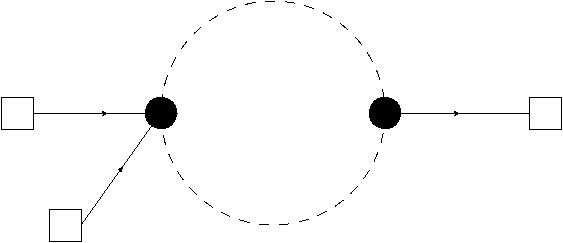}
 \end{center}
 \caption{Leading graph of the inflaton three point function in the absence of spontaneous symmetry breaking, 
  which is inevitably becomes at one-loop level. 
  The rigid line represents the inflaton, and the dotted line represents the Higgs boson in GUT. 
  See Appendix B for notation.}
 \label{Fig1}
 \end{figure}

\begin{figure}[htbp]
 \begin{center}
  \includegraphics[width=15cm]{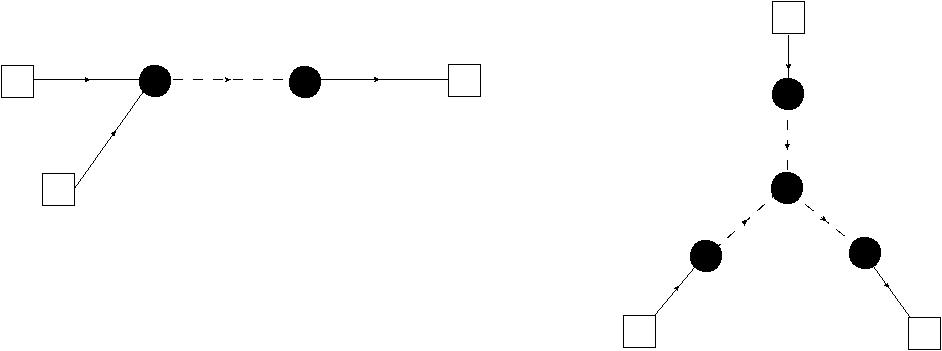}
 \end{center}
 \caption{Leading graph of the inflaton three point function in the presence of spontaneous symmetry breaking. 
 In the absence of spontaneous symmetry breaking, no such tree graph exists.}
 \label{Fig2}
 \end{figure}

Non-Gaussianity is the three- or higher-point function of some quantum fluctuation in the curvature perturbations.
Three point functions in models with only inflatons and gravitons were computed by Maldacena \cite{Maldacena}. 
The effective field theory of inflation was proposed by C. Cheung, {\it et al}. \cite{Cheung}, 
 and its formalism has been used to calculate three-point functions in models with various particles. 
We focus on the case where the GUT scale is close to the energy scale of inflation 
 and calculate three point function of inflaton by exchanging Higgs boson in GUT. 
The characteristic feature of this model is that the interaction between Higgs boson and inflaton 
 generated by the (non-)vanishing Higgs boson vacuum expectation value (VEV) 
 contributes to three point function of the inflaton at the (tree) 1-loop level 
 as shown in Fig.\ref{Fig1} and Fig.\ref{Fig2}. 
The results are found to be consistent with the current observational constraints on non-Gaussianity 
 without drastic fine tuning of parameters, 
and it might be possible to detect the signature of the Higgs boson in GUT 
 by 21cm spectrum, future LSS and future CMB depending on our model parameters. 

This paper is organized as follows. 
First, we discuss the setup of our model. 
The effective field theory of inflation is briefly introduced, giving the propagators of the inflaton as NG boson $\pi$. 
Furthermore, we introduce the Higgs potential in GUT 
 and confirm that the Higgs boson interacts with the inflaton linearly after developing the vacuum expectation value 
 due to the spontaneous symmetry breaking. 
In Section 3, we actually calculate the non-Gaussianity. 
Concretely, we compute 
 three point function of the inflatons via the Higgs boson exchange at tree level 
 by using the approximation in horizon exit. 
Non-Gaussianity is evaluated from the obtained three-point functions and 
 the results are compared with the observational data by the Planck satellite. 
Conclusions are given in Section 4. 
In addition, Appendix A presents calculations on determinants of the metric, and Appendix B summarizes in-in formalism \cite{in-in}.

\section{A Model}
We consider the Friedmann-Lemaitre-Robertson-Walker (FLRW) spacetime with fluctuations 
 and curvature $K=0$ expressed in the ADM formalism as the inflationary spacetime.
\begin{equation}
ds^{2} = -N^{2}dt^{2} + \tilde{h}_{ij} \left( dx^{i} +N^{i}dt  \right) \left( dx^{j} +N^{j}dt  \right), 
\end{equation}
where $\tilde{h}_{ij}$ is a spatial components of the metric, 
 $N$ is the lapse function and $N^{i}$ is the shift function.  
The tilde represents a physical quantity in the comoving gauge. The action we consider is
\begin{equation}
S = S_{\mathrm{grav}} + S_{\text{SU(5) gauge}}  + S_{\text{SU(5) Higgs}} ,
\end{equation}
where $S_{\mathrm{grav}}, S_{\text{SU(5) gauge}}$ and $S_{\text{SU(5) Higgs}}$ describe 
the actions of the gravity, SU(5) gauge bosons 
and 
an SU(5) adjoint Higgs boson to break GUT gauge symmetry 
 respectively\footnote{In the following, we consider in this paper the SU(5) GUT as an illustrating example, 
 but it can be easily extended to other GUT gauge group.}.
We will now examine each action in detail 
but omit the $S_{\text{SU(5) gauge}}$ since it is unnecessary for our computation of the graph Fig.\ref{Fig2}. 
From the viewpoint of the effective field theory \cite{Cheung, Leonardo}, 
 the Einstein-Hilbert action and the inflaton action after the transformation of time coordinates 
 $t \mapsto \tilde{t} = t - \pi(\tilde{t}, \bm{x})$ can be written as
\begin{equation}
\begin{aligned} 
S_{\mathrm{grav}} = \int d^{4} x \sqrt{-g} 
 &\left[\frac{1}{2} M_{\mathrm{Pl}}^{2} R - M_{\mathrm{Pl}}^{2} \left(3 H^{2}(\tilde{t}+\pi)
 +\dot{H}( \tilde{t}+\pi)\right) \right. \\
 & \left. +M_{\mathrm{PI}}^{2} \dot{H}( \tilde{t}+\pi)\left(\partial_{\mu}( \tilde{t}+\pi) \partial_{\nu}
 ( \tilde{t}+\pi) g^{\mu \nu}\right) \right. \\
 & +\frac{M_{2}( \tilde{t}+\pi)^{4}}{2 !}\left(\partial_{\mu}( \tilde{t}+\pi) \partial_{\nu}( \tilde{t}+\pi) 
 g^{\mu \nu}+1\right)^{2}\\
 &+ \left.\frac{M_{3}( \tilde{t}+\pi)^{4}}{3 !}\left(\partial_{\mu}( \tilde{t}+\pi) \partial_{\nu}( \tilde{t}+\pi) 
 g^{\mu \nu}+1\right)^{3}+\ldots
 \right], 
 \end{aligned}
\end{equation}
where $g$ is the determinant of the metric $g_{\mu\nu}$, 
 $M_{\mathrm{Pl}}$ is the Planck mass, 
 $R$ is the Ricci curvature in four dimensions, $H$ is the Hubble parameter, 
 $\pi$ is a Nambu-Goldstone boson of time translation which is identified with the inflaton.  
$M_{2, 3}$ are the coefficients of the high-dimensional operators. 

The quadratic terms of the effective action of inflaton 
$\pi$ is identified as \cite{Cheung, Mukohyama}
\begin{equation}
I_{2} = M_{\text{Pl}}^{2} \int dt d^{3}x\, a^{3} \left[ -\frac{\dot{H}}{c_{s}^{2}}  
\left( \dot{\pi}^{2} - c^{2}_{s} \frac{ \left( \partial_{i} \pi \right)^{2}}{ a^{2}}  \right) \right].
\end{equation}
The sound speed $c_{s}$ is a quantity that is not determined by the effective field theory, 
 but is bounded by the fundamental theory and observational data. 
The mode function $w(\tau, k)$ of the inflaton
\ $\pi$ is obtained as
\begin{equation}
w(\tau, k) = \frac{c_{s}}{\sqrt{2\epsilon}a H M_{\text{Pl}}} \frac{1+ ic_{s} k \tau } { \sqrt{2c_{s}k} c_{s}k\tau } e^{-ic_{s}k\tau},
\end{equation}
where $\tau$ is the conformal time and $\epsilon$ is the slow roll parameter. 
Using the expression for the propagators by in-in formalism in Appendix B, 
 the propagators of inflaton 
 $\pi$ is obtained as
\begin{eqnarray}
\Delta_{>}\left(\tau_{1}, \tau_{2}, k\right) &=& w\left(\tau_{1}, k\right) w^{*}\left(\tau_{2}, k\right) \nonumber \\
&=& \frac{ 1+ ic_{s}k\left(  \tau_{1} - \tau_{2} \right) 
+ (c_{s}k)^{2} \tau_{1} \tau_{2} }   { 4 \epsilon M_{\text{Pl}} c_{s} k^{3} } e^{-i c_{s} k ( \tau_{1} - \tau_{2})} ,  \label{Delta>}  \\
\Delta_{<}\left(\tau_{1}, \tau_{2}, k\right) &=& w\left(\tau_{2}, k\right) w^{*}\left(\tau_{1}, k\right) \nonumber \\
&=& \frac{ 1- ic_{s}k\left(  \tau_{1} - \tau_{2} \right) 
+ (c_{s}k)^{2} \tau_{1} \tau_{2} }   { 4 \epsilon M_{\text{Pl}} c_{s} k^{3} } e^{i c_{s} k ( \tau_{1} - \tau_{2})} \label{Delta<}
\end{eqnarray}

Next, we discuss the action of the Higgs boson for GUT gauge symmetry breaking to the SM gauge symmetry. 
Let us denote $\Sigma'$ for the Higgs of the 24-dimensional adjoint representation of SU(5) 
 and its renormalizable potential is introduced as  
\begin{equation}
V(\Sigma') = - M^{2} \text{tr} (\Sigma'^{2}) + \lambda_{1} \left\{\text{tr} (\Sigma'^{2}) \right\}^{2} 
+  \lambda_{2}\,  \text{tr} (\Sigma'^{4}), 
\label{Higgs potential}
\end{equation}
where $M, \lambda_{1}$ and $\lambda_{2}$ are constants, and $M$ is of order of the GUT scale. 
\begin{equation}
M \sim \mathcal{O} (M_{\text{GUT}}).
\end{equation}
 We expand the adjoint Higgs $\Sigma'$ around the expectation value
\begin{equation}
\langle \Sigma \rangle = v \  \text{diag} (2, 2, 2, -3, -3), \quad v \sim \mathcal{O} (M_{\text{GUT}})
\end{equation}
as
\begin{equation}
\Sigma' = \langle \Sigma \rangle + \Sigma. 
\end{equation}
Now, we calculate three point function of inflaton 
to extract non-Gaussianity by the existence of $\Sigma$, 
which is given by a graph of tree level exchange of $\Sigma$ shown in Fig. \ref{Fig2}. 
In order to extract the first-order term of $\Sigma$ in the potential, 
we expand $\Sigma'^{2}$ around the expectation value yields
\begin{eqnarray}
\Sigma'^{2} &=& \left(  \langle \Sigma \rangle + \Sigma \right)^{2} \nonumber \\
&=& \langle \Sigma \rangle ^{2} + \langle \Sigma \rangle \Sigma + \Sigma \langle \Sigma \rangle + \Sigma^{2}, \label{Higgs 2}
\end{eqnarray}
and we obtain 
\begin{eqnarray}
2 \text{tr} \left(\langle \Sigma \rangle \Sigma\right) &=& 2 \times 
\left(  2v \Sigma_{11}  +  2v \Sigma_{22}  +  2v \Sigma_{33}  -  3v \Sigma_{44}  -  3v \Sigma_{55}  \right) \nonumber \\
&=&  2 \times 5v \left(   \Sigma_{11}  +   \Sigma_{22}  +   \Sigma_{33}  \right) \nonumber \\
&=& 10 v \left(   \Sigma_{11}  +   \Sigma_{22}  +   \Sigma_{33}  \right)
\end{eqnarray}
from the terms in $\text{tr} (\Sigma'^{2})$. 
Note that 
only diagonal components of $\Sigma$ are taken into account and 
the traceless condition for $\Sigma$
\begin{equation}
 \Sigma_{11}  +   \Sigma_{22}  +   \Sigma_{33} + \Sigma_{44}  +   \Sigma_{55} = 0 . \label{traceless}
\end{equation}
is used in the second equality. 
Then, since the first term of (\ref{Higgs 2}) can be computed as
\begin{equation}
\text{tr}\left( \langle \Sigma \rangle^{2} \right) = 30 v^{2},
\end{equation}
we have
\begin{equation}
 \text{tr} (\Sigma'^{2}) = 30v^{2} +10 v \left(   \Sigma_{11}  +   \Sigma_{22}  +   \Sigma_{33}  \right) +  \text{tr} (\Sigma^{2}).
\end{equation}
Now we can obtain
\begin{eqnarray}
\lambda_{1} \left\{\text{tr} (\Sigma'^{2}) \right\}^{2} &=& \lambda_{1}  \left[30v^{2} 
+ 10 v \left(   \Sigma_{11}  +   \Sigma_{22}  +   \Sigma_{33}  \right) +  \text{tr} (\Sigma^{2}) \right]^{2} \nonumber \\
&\supset&  600 \lambda_{1} v^{3}  \left(   \Sigma_{11}  +   \Sigma_{22}  +   \Sigma_{33}  \right)
\end{eqnarray}
from the second term in potential (\ref{Higgs potential}). 
Similarly, 
\begin{equation}
\lambda_{2} \, \text{tr} (\Sigma'^{4})  \supset  140 \lambda_{2} v^{3}  \left(   \Sigma_{11}  +   \Sigma_{22}  +   \Sigma_{33}  \right)
\end{equation}
is obtained from the third term in (\ref{Higgs potential}).\\
\ Next, we consider third-order term for $\Sigma$ in potential (\ref{Higgs potential}):
\begin{eqnarray}
\lambda_{1} \left\{\text{tr} (\Sigma'^{2}) \right\}^{2} &=& \lambda_{1}  \left[30v^{2} 
+ 10 v \left(   \Sigma_{11}  +   \Sigma_{22}  +   \Sigma_{33}  \right) +  \text{tr} (\Sigma^{2}) \right]^{2} \nonumber \\
&\supset&  20 \lambda_{1} v  \left(   \Sigma_{11}  +   \Sigma_{22}  +   \Sigma_{33}  \right)  \text{tr} (\Sigma^{2})
\end{eqnarray}
The last term $\text{tr} (\Sigma^{2})$ is
\begin{equation}
\text{tr} (\Sigma^{2}) = \Sigma^{2}_{11} + \Sigma^{2}_{22} + \cdots 
+ \Sigma^{2}_{55} + 2( \Sigma_{12}\Sigma_{21} + \Sigma_{13}\Sigma_{31} + \cdots  + \Sigma_{45}\Sigma_{54}),
\end{equation}
but the only meaningful term are only $\Sigma^{2}_{11} + \Sigma^{2}_{22} + \cdots + \Sigma^{2}_{55}$, 
since the $\pi \Sigma$ term has only a diagonal component $\Sigma_{11}, \Sigma_{22}, \Sigma_{33}$ 
when performing the graph calculation. $\Sigma^{2}_{44}  +   \Sigma^{2}_{55}$ can be rewritten as
\begin{equation}
 \Sigma^{2}_{44}  +   \Sigma^{2}_{55} = \left(   \Sigma_{11}  +   \Sigma_{22}  +   \Sigma_{33}  \right)^{2} - 2\Sigma_{44} \Sigma_{55} 
\end{equation}
using (\ref{traceless}). Hence, the meaningful part is \begin{eqnarray}
\text{tr} (\Sigma^{2}) &\supset& \Sigma^{2}_{11} + \Sigma^{2}_{22} + \Sigma^{2}_{33} + \Sigma^{2}_{44} + \Sigma^{2}_{55} \nonumber \\
&\supset& \Sigma^{2}_{11} + \Sigma^{2}_{22} + \Sigma^{2}_{33} +\left(   \Sigma_{11}  +   \Sigma_{22}  +   \Sigma_{33}  \right)^{2} \nonumber \\
&=& 2 \left( \Sigma^{2}_{11} + \Sigma^{2}_{22} + \Sigma^{2}_{33}  \right) + 2 \left(   \Sigma_{11} \Sigma_{22}  +   \Sigma_{22} \Sigma_{33}  +   \Sigma_{33}  \Sigma_{11} \right)
\end{eqnarray}
and
\begin{eqnarray}
\lambda_{1} \left\{\text{tr} (\Sigma'^{2}) \right\}^{2} &\supset& 40 \lambda_{1} v \left( \Sigma^{3}_{11} + \Sigma_{22}^{3}+ \Sigma_{33}^{3} \right. \nonumber \\
&&\left. + 2\Sigma_{11}^{2}\Sigma_{22} + 2\Sigma_{11}^{2}\Sigma_{33} + 2\Sigma_{22}^{2}\Sigma_{11} + 2\Sigma_{22}^{2}\Sigma_{33} + 2\Sigma_{33}^{2}\Sigma_{22} + 2\Sigma_{33}^{2}\Sigma_{11} \right. 
\nonumber \\
&&\left. + 3\Sigma_{11}\Sigma_{22}\Sigma_{33} \right)
\end{eqnarray}
is obtained. 
Although there are cross terms among diagonal components, their contributions to non-Gaussianity are comparable, 
therefore we only need to consider $40 \lambda_{1} v \Sigma^{3}_{11}$ in effect. 
We also omitted $\Sigma_{44} \Sigma_{55}$, since they are functions of $\Sigma_{11}, \Sigma_{22}$ 
and $\Sigma_{33}$ and they give the same order of contribution as the cross-terms. 
Furthermore, the third-order term of $\Sigma$ is also obtained 
from the term $\lambda_{2}\,  \text{tr} (\Sigma'^{4})$ of potential (\ref{Higgs potential}), 
since it only yields the same order of contribution, we write them collectively as $\alpha$ later.
The necessary part of action for the adjoint Higgs boson of SU(5)
\begin{equation}
S_{\text{SU(5) Higgs}} = \int d^{4}x\, \sqrt{-g} \left[  \left( D_{\mu} \Sigma \right)^{\dagger}  D^{\mu} \Sigma - V(\Sigma)\right]
\end{equation}
for calculation in Fig.\ref {Fig2} 
is
\begin{eqnarray}
S_{\text{SU(5) Higgs}} &\supset& \int d^{4}x\, \sqrt{-g} 
\left[  \left( \partial_{\mu} \Sigma \right)^{\dagger}  \partial^{\mu} \Sigma +  10 M^{2} v 
\left(   \Sigma_{11}  +   \Sigma_{22}  +   \Sigma_{33}  \right) \right. \nonumber \\
&& - \left.  \left(600 \lambda_{1} + 140 \lambda_{2}\right) v^{3}  
\left(   \Sigma_{11}  +   \Sigma_{22}  +   \Sigma_{33}  \right) + 40 \lambda_{1} v 
\left( \Sigma^{3}_{11} + \Sigma_{22}^{3}+ \Sigma_{33}^{3} \right)\right]. \nonumber \\
\end{eqnarray}
To calculate the graph in Fig.\ref{Fig2}, 
we have to extract the terms proportional to $\pi$ and $\pi^2$ from the metric $\sqrt{-g}$. 
From appendix A, the metric $\sqrt{-g}$ is
\begin{equation}
\sqrt{-g} = a^{3}(\tilde{t}) + 3a^{3} H 
\pi - a^{3} \dot{\pi} 
+ \frac{2}{9} a^{3} H^{2} 
\pi^{2} 
- 3a^{3}H 
\pi \dot{\pi} + a^{3} \dot{\pi}^{2}
\end{equation}
after the transformation
\begin{equation}
t \mapsto \tilde{t} = t - \pi(\tilde{t}, \bm{x})
\end{equation}
of time coordinates. Thus, the interaction with the NG boson $\pi$ with no derivative is
\begin{eqnarray}
&& \sqrt{-g}   \left[   10 M^{2} v \left(   \Sigma_{11}(t)  +   \Sigma_{22}(t)  +   \Sigma_{33}(t)  \right) -  \left(600 \lambda_{1} + 140 \lambda_{2}\right)v^{3}  \left(   \Sigma_{11} (t) +   \Sigma_{22}(t)  +   \Sigma_{33}(t)  \right) \right] \nonumber \\
&=&  \sqrt{-g} \left(   10M^{2} v -  \left(600 \lambda_{1} + 140 \lambda_{2}\right) v^{3} \right)  \left(   \Sigma_{11}(t)  +   \Sigma_{22} (t) +   \Sigma_{33}(t)  \right) \nonumber \\
&=&  3  \left(  10M^{2} v -  \left(600 \lambda_{1} + 140 \lambda_{2}\right) v^{3} \right)a^{3} H \left(  1+ \frac{\epsilon}{3}\right) \pi   \left(   \Sigma_{11}(\tilde{t})  +   \Sigma_{22}(\tilde{t})   +   \Sigma_{33}(\tilde{t})    \right) \nonumber \\
&& + \frac{2}{9}  \left(  10M^{2} v - \left(600 \lambda_{1} + 140 \lambda_{2}\right)v^{3} \right) a^{3} H^{2} 
\left( 1 + \frac{1}{3} \epsilon \right) \pi^{2}   \left(   \Sigma_{11}(\tilde{t})    
+   \Sigma_{22}(\tilde{t})   +   \Sigma_{33}(\tilde{t})    \right) \nonumber \\
&& + 3  \left(  10M^{2} v -  \left(600 \lambda_{1} + 140 \lambda_{2}\right) v^{3} \right)a^{3} H 
\left(  1+ \frac{\epsilon}{3}\right) \pi^{2}   \left(   \dot{\Sigma}_{11}(\tilde{t})  
+  \dot{\Sigma}_{22}(\tilde{t})   +  \dot{\Sigma}_{33}(\tilde{t})    \right), 
\nonumber \\
&&
\end{eqnarray}
and with time derivative is
\begin{eqnarray}
&&\sqrt{-g} \left[    10 M^{2} v  - \left( 600\lambda_{1} + 140\lambda_{2} \right) v^{3}
 \right] \left(   \Sigma_{11}(t)  +   \Sigma_{22}(t)   +   \Sigma_{33}(t)    \right) \nonumber \\
 &=&    -a^{3} \left\{     10 M^{2} v  - \left( 600\lambda_{1} + 140\lambda_{2} \right) v^{3}
 \right\} \dot{\pi} \left(   \Sigma_{11}(\tilde{t})  
+  \Sigma_{22}(\tilde{t})   +  \Sigma_{33}(\tilde{t})    \right) \nonumber \\
&& - a^{3}H \left( 3+ 2\epsilon  \right) \left\{     10 M^{2} v  - \left( 600\lambda_{1} + 140\lambda_{2} \right) v^{3}
 \right\} \pi \dot{\pi} \left(   \Sigma_{11}(\tilde{t})  
+  \Sigma_{22}(\tilde{t})   +  \Sigma_{33}(\tilde{t})    \right) \nonumber \\
 &&  -a^{3} \left\{     10 M^{2} v  - \left( 600\lambda_{1} + 140\lambda_{2} \right) v^{3}
 \right\} \pi \dot{\pi} \left(   \dot{\Sigma}_{11}(\tilde{t})  
+  \dot{\Sigma}_{22}(\tilde{t})   +  \dot{\Sigma}_{33}(\tilde{t})    \right)   \nonumber \\
 &&  +a^{3} \left\{     10 M^{2} v  - \left( 600\lambda_{1} + 140\lambda_{2} \right) v^{3}
 \right\}\dot{\pi}^{2} \left(  \Sigma_{11}(\tilde{t})  
+  \Sigma_{22}(\tilde{t})   +  \Sigma_{33}(\tilde{t})    \right)   
\end{eqnarray}
Since $\Sigma_{11}, \Sigma_{22}$ and $\Sigma_{33}$ are equivalent, we consider $\Sigma_{11}$ in the following. 

Next, we consider the propagator of $\Sigma$, since the interaction has been obtained. 
The second order term of potential (\ref{Higgs potential}) is
\begin{eqnarray}
&& - M^{2} \text{tr} (\Sigma'^{2}) \supset -M^{2} \left(\Sigma_{11}\right)^{2},  \\
&&  \lambda_{1} \left\{\text{tr} (\Sigma'^{2}) \right\}^{2} \supset 160 \lambda_{1} v^{2}  \left(\Sigma_{11}\right)^{2},  \\
&&  \lambda_{2} \text{tr} (\Sigma'^{4}) \supset 24 \lambda_{2} v^{2} \left(\Sigma_{11}\right)^{2}.
\end{eqnarray}
Hence, by setting
\begin{equation}
-m^{2} := -M^{2} + 160\lambda_{1} v^{2} +  24\lambda_{2} v^{2} ,
\end{equation}
$\Sigma_{11}$ is a scalar field with mass $m^{2}$ in de Sitter spacetime, therefore it has propagator
\begin{eqnarray}
G_{>}\left(\tau_{1}, \tau_{2}, k\right) &=& - i \frac{\sqrt{\pi}}{2} e^{i\pi \left( \nu/2 + 1/4 \right)} H 
\left(  -\tau_{1} \right) ^{3/2} H^{(1)}_{\nu}(-k\tau_{1}) \times  \nonumber \\
&&i \frac{\sqrt{\pi}}{2} e^{i\pi \left( \nu^{*}/2 + 1/4 \right)} H \left(  -\tau_{2} \right) ^{3/2} H^{(2)}_{\nu}(-k\tau_{2}) \nonumber \\
&=& -\frac{ \pi }{ 4} e^{-\pi \text{Im}(\nu)}  H^{2} \left(  \tau_{1} \tau_{2} \right) ^{3/2} H^{(1)}_{\nu}(-k\tau_{1}) H^{(2)}_{\nu}(-k\tau_{2}), \label{G>} \\
G_{<}\left(\tau_{1}, \tau_{2}, k\right) &=& -\frac{ \pi }{ 4} e^{-\pi \text{Im}(\nu)}  H^{2} \left(  \tau_{1} \tau_{2} \right) ^{3/2} H^{(1)}_{\nu}(-k\tau_{2}) H^{(2)}_{\nu}(-k\tau_{1}). \label{G<}
\end{eqnarray}
from appendix B. From the viewpoint of effective field theory, since $m^{2}$ should be around the Hubble scale, $\lambda_{1}$ and $\lambda_{2}$ must be an order of $\mathcal{O} (10^{-2})$.

\section{Calculation of non-Gaussianity}
\begin{figure}[htbp]
 \begin{center}
  \includegraphics[width=14cm]{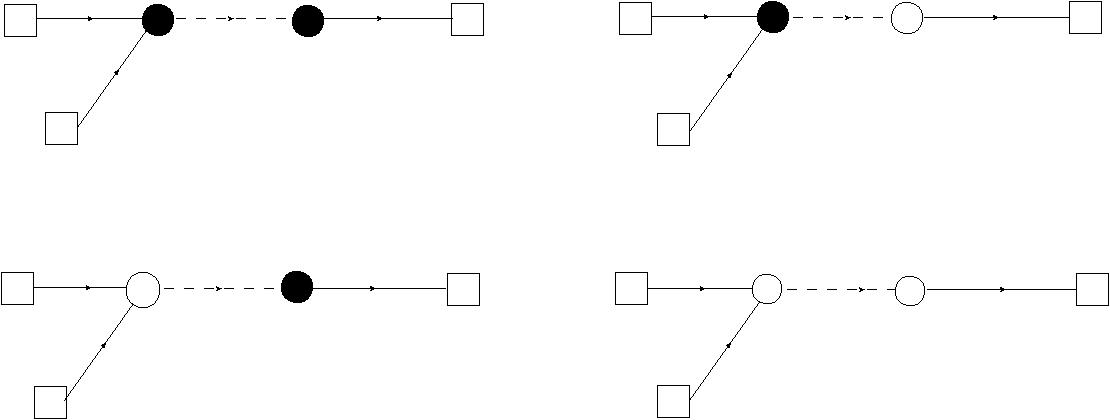}
 \end{center}
 \caption{Four graphs in in-in formalism. Black circles indicate $+$ and white circles indicate $-$.}
 \label{Fig3}
 \end{figure}

We are now ready to calculate non-Gaussianity. 
Since the interaction of $\Sigma_{11}$ and the interaction of $\dot{\Sigma}_{11}$ give contributions of the same order to non-Gaussianity, 
 only $\Sigma_{11}$ is considered and the latter is doubled. First, consider interactions that have no time derivative. 
According to the rule of in-in formalism, 
 there exist four types of graphs, $(+, +)$, $(+, -)$, $(-, +)$ and $(-, -)$ as shown in Fig.\ref{Fig3}, 
 mediated by one $\Sigma$ between the three inflatons 
 $\pi$. 

\begin{figure}[htbp]
 \begin{center}
  \includegraphics[width=14cm]{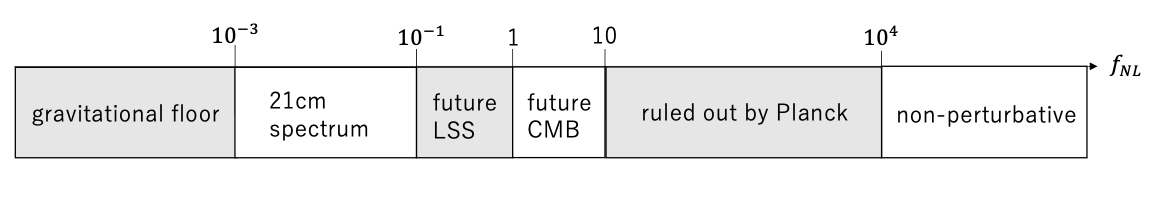}
 \end{center}
 \caption{Schematic illustration of current 
 and future constraints on the non-Gaussianity 
 (Figure taken from \cite{Baumann2018}).}
 \label{Fig4}
 \end{figure}

As an example, let us consider the graphs of $(+, -)$ and $(-, +)$. The sum of the two graphs is given as follows:
\begin{eqnarray}
\langle \pi \pi \pi \rangle &=& \int_{-(1+ i\epsilon) \infty}^{0} a(\tau_{1})d\tau_{1}\,\int_{-(1+ i\epsilon) \infty}^{0} 
a(\tau_{2})d\tau_{2} (-i)   \frac{2}{9}  \left( 10M^{2} v -  \left(600 \lambda_{1} + 140 \lambda_{2}\right) v^{3} \right) \nonumber \\
&& \times a^{3}(\tau_{1}) H^{2} 
\times  3 i  \left( 10M^{2} v -  \left(600 \lambda_{1} + 140 \lambda_{2}\right)v^{3} \right)a^{3}(\tau_{2}) H 
\nonumber \\
&& \times G_{+-}(\tau_{1}, \tau_{2}, p_{3}) 
\Delta_{+-}(0, \tau_{1}, p_{1})  \Delta_{+-}(0, \tau_{1}, p_{2})  \Delta_{++}(\tau_{2}, 0, p_{3}) + (p_{1}\leftrightarrow p_{3})  \label{pi3}
\\
&& + (p_{2}\leftrightarrow p_{3}) +\text{c.c.} \nonumber \\
&=& \frac{2}{3} \left( 10M^{2} v -  \left(600 \lambda_{1} + 140 \lambda_{2}\right) v^{3} \right)^{2} H^{3} 
\nonumber \\
&& \times \int_{-(1+ i\epsilon) \infty}^{0} d\tau_{1}\,\int_{-(1+ i\epsilon) \infty}^{0} d\tau_{2}\, a^{4} (\tau_{1}) a^{4} (\tau_{2}) 
G_{+-}(\tau_{1}, \tau_{2}, p_{3})  \Delta_{+-}(0, \tau_{1}, p_{1})   \nonumber \\
&& \times \Delta_{+-}(0, \tau_{1}, p_{2})  \Delta_{++}(\tau_{2}, 0, p_{3}) + (p_{1}\leftrightarrow p_{3}) + (p_{2}\leftrightarrow p_{3}) +\text{c.c.} 
\end{eqnarray}
Note that we neglect the terms including $\epsilon$ which are discussed in Appendix A in three point functions of inflaton, 
 since they make only sub-leading contributions to non-Gaussianity. 
 Now, substituting the propagator (\ref{G<}) of $\Sigma$ and the propagators (\ref{Delta>}) and (\ref{Delta<}) of $\pi$, we have
 \begin{eqnarray}
\langle \pi \pi \pi \rangle 
&=& - \frac{\pi }{2^{7} \cdot 3}  e^{-\pi \text{Im}(\nu)}   
\left( 10M^{2} v -  \left(600 \lambda_{1} + 140 \lambda_{2}\right) v^{3} \right)^{2}  
\frac{1}   { \left( H \epsilon M_{\text{Pl}} c_{s} p_{1}p_{2}p_{3}\right)^{3}} \nonumber \\
&& \times  \int_{-(1+ i\epsilon) \infty}^{0} d\tau_{1}\,
\int_{-(1+ i\epsilon) \infty}^{0} d\tau_{2}\,  
\left(  \tau_{1} \tau_{2} \right)^{-5/2} H^{(1)}_{\nu}(-p_{3}\tau_{1}) H^{(2)}_{\nu}(-p_{3}\tau_{2}) \nonumber \\
&&  \times  \left(1+ ic_{s}p_{1}  \tau_{1} \right)    \left( 1+ ic_{s}p_{2}  \tau_{1} \right) 
\left( 1+ ic_{s}p_{3}  \tau_{2} \right)  e^{-i c_{s} \left(p_{1} + p_{2} \right)  \tau_{1} } e^{-i c_{s} p_{3}  \tau_{2} } \nonumber \\
&&+ (p_{1}\leftrightarrow p_{3}) + (p_{2}\leftrightarrow p_{3}) +\text{c.c.} \label{3pi1}
\end{eqnarray}
As discussed in \cite{ChenWang2} and \cite{BaumannGreen}, 
 we use the approximation of the Hankel function
\begin{eqnarray}
&& H^{(1)}_{\nu}(-p_{3}\tau_{2}) \to - i \frac{ 2^{\nu}} { \pi} (-p_{3}\tau_{2})^{-\nu} \Gamma(\nu), \label{Hankel approximation1}\\
&& H^{(2)}_{\nu}(-p_{3}\tau_{1}) \to  i \frac{ 2^{\nu}} { \pi} (-p_{3}\tau_{1})^{-\nu} \Gamma(\nu) \label{Hankel approximation2}
\end{eqnarray}
with horizon exit $-p_{3}\tau_{1}, -p_{3}\tau_{2}\to 1$ to evaluate the integral. 
This is an approximation to extract the effect that the contribution is the largest as time evolves and the fluctuations freeze. 
Note that we consider the region
\begin{equation}
0< \nu \leq \frac{3}{2}, 
\end{equation}
in other words, 
\begin{equation}
0 < \frac{m^{2}}{H^{2}} \leq \frac{9}{4}
\end{equation}
where the suppression factor $e^{-\pi\text{Im}(\nu)}$ does not appear.
Substituting approximations (\ref{Hankel approximation1}) and (\ref{Hankel approximation2}) of the Hankel functions into (\ref{3pi1}), 
 three point function of inflaton $\langle \pi \pi \pi \rangle $ becomes
\begin{eqnarray}
\langle \pi \pi \pi \rangle 
&=&   - \frac{ \Gamma^{2}(\nu) }{2^{7-2\nu} \cdot 3 \pi}    
\left( 10M^{2} v -  \left(600 \lambda_{1} + 140 \lambda_{2}\right) v^{3} \right)^{2}  
\frac{1}{ \left( H \epsilon M_{\text{Pl}} c_{s} p_{1}p_{2}p_{3}\right)^{3}} 
\frac{1}{p_{3}^{2\nu}} \nonumber \\
&& \times  \int_{-(1+ i\epsilon) \infty}^{0} d\tau_{1}\,\int_{-(1+ i\epsilon) \infty}^{0} d\tau_{2}\,    
\left(  \tau_{1} \tau_{2} \right) ^{-5/2-\nu} 
e^{-i c_{s} \left(p_{1} + p_{2} \right)  \tau_{1} } e^{-i c_{s} p_{3}  \tau_{2} } \nonumber \\
&& \times 
\left(1+ ic_{s}p_{1}  \tau_{1} \right)    
\left( 1+ ic_{s}p_{2}  \tau_{1} \right) \left( 1+ ic_{s}p_{3}  \tau_{2} \right) 
+ (p_{1}\leftrightarrow p_{3}) + (p_{2}\leftrightarrow p_{3}) +\text{c.c.} \label{3pi2} \nonumber \\
\end{eqnarray}
Note that the dependence on the external momentum $p_3$ is found to be non-local. 
This implies that the contribution for the three point function actually comes from the $\Sigma$ field. 
Non-Gaussianity is defined by
\begin{equation}
\langle \zeta \zeta \zeta \rangle = (2\pi)^{7} \delta^{(3)} 
( \bm{p}_{1} +  \bm{p}_{2} +  \bm{p}_{3})\tilde{p}^{2}_{\zeta} \left( \frac{9}{10} f_{\text{NL}} \right) \frac{1}{(p_{1}p_{2}p_{3})^{2}} 
\end{equation}
in the case where the configuration of the external momentum are equilateral, 
\begin{equation}
p:= p_{1} =p_{2} =p_{3}
\end{equation}
in the following. 
Note that $\zeta$ is the curvature fluctuation, which is related to inflaton 
$\pi$ by $\zeta = -H \pi$. 
Since the integral only has an effect up to the horizon exit $-p_{3}\tau_{1}, -p_{3}\tau_{2}\to 1$, 
 the upper limit of the integral should be \cite{BaumannGreen}
\begin{equation}
\tau_{1*} = \tau_{2*}= -p^{-1}.
\end{equation}
In this case, the higher order term of the factor $e^{-i c_{s} \left(p_{1} + p_{2} \right) \tau_{1} } e^{-i c_{s} p_{3} \tau_{2} }$ 
 is a small quantity, then the integral of \label{3pi2} can be calculated as follows:
\begin{eqnarray}
&& \int_{-(1+ i\epsilon) \infty}^{0} d\tau_{1}\,\int_{-(1+ i\epsilon) \infty}^{0} d\tau_{2}\, 
\left(  \tau_{1} \tau_{2} \right) ^{-5/2-\nu} \left(1+ ic_{s}p_{1}  \tau_{1} \right)    \left( 1+ ic_{s}p_{2}  \tau_{1} \right) 
\left( 1+ ic_{s}p_{3}  \tau_{2} \right)  
\nonumber \\
&&\times e^{-i c_{s} \left(p_{1} + p_{2} \right)  \tau_{1} } e^{-i c_{s} p_{3}  \tau_{2} } \nonumber \\
&=&  \left( \frac{-2}{3+2\nu}  - \frac{4ic_{s}} {1+2\nu}\right) \left( \frac{-2}{3+2\nu}  - \frac{2ic_{s}} {1+2\nu}\right) p^{3 + 2\nu}.
\end{eqnarray}
Therefore, the three point function of inflaton 
$\pi$ can be computed as
\begin{eqnarray}
\langle \pi \pi \pi \rangle 
&\simeq &     - \frac{ \Gamma^{2}(\nu) }{2^{7-2\nu} \cdot 3 \pi}     
\left( 10M^{2} v -  \left(600 \lambda_{1} + 140 \lambda_{2}\right) v^{3} \right)^{2}  
\frac{1}   { \left( H \epsilon M_{\text{Pl}} c_{s} p_{1}p_{2}p_{3}\right)^{3}} 
\frac{1}{p_{3}^{2\nu}} \nonumber \\
&& \times   \left( \frac{-2}{3+2\nu}  - \frac{4ic_{s}} {1+2\nu}\right) \left( \frac{-2}{3+2\nu} 
 - \frac{2ic_{s}} {1+2\nu}\right) p^{3 + 2\nu}  + (p_{1}\leftrightarrow p_{3}) + (p_{2}\leftrightarrow p_{3}) +\text{c.c.} \nonumber \\
&=&      - \frac{ \Gamma^{2}(\nu) }{2^{7-2\nu} \cdot 3 \pi}     
\left( 10M^{2} v + \left(600 \lambda_{1} + 140 \lambda_{2}\right)v^{3} \right)^{2}  
\nonumber \\
&& \times  \frac{1}   { \left( H \epsilon M_{\text{Pl}} c_{s} \right)^{3}} p^{-6} 
\left( \frac{8}{(3+2\nu)^{2}} -   \frac{16c^{2}_{s}}{(1+ 2\nu)^{2}}  \right) \times 3 
\end{eqnarray}
Since the relation between the curvature fluctuation $\zeta$ and inflaton 
$\pi$ is $\zeta = -H \pi$, 
the non-Gaussianity can be estimated as
\begin{eqnarray}
f_{\text{NL}} 
&=&  \frac{10}{9} \frac{1}{(2\pi)^{7}\tilde{p}^{2}_{\zeta}} p^{6} M^{-3}_{\text{Pl}}  (-H)^{3} \nonumber \\
&& \times       \frac{ - \Gamma^{2}(\nu) }{2^{7-2\nu} \cdot 3 \pi}     
\left( 10M^{2} v -  \left(600 \lambda_{1} + 140 \lambda_{2}\right) v^{3} \right)^{2}  \nonumber \\
&& \times \frac{1}   { \left( H \epsilon M_{\text{Pl}} c_{s} \right)^{3}} p^{-6} \left( \frac{8}{(3+2\nu)^{2}} 
-   \frac{16c^{2}_{s}}{(1+ 2\nu)^{2}}  \right) \times 3   \label{f NL 1} 
\end{eqnarray}
by reviving $M_{\text{Pl}}$ for $\zeta$. Here, the constant $c_{s}$ has the relation $r=16\epsilon c_{s}$ 
 with the tensor-scalar ratio $r$ and the upper bound $r<0.036$ \cite{BICEP/Keck} gives 
\begin{equation}
\epsilon c_{s} \sim \mathcal{O} (10^{-3 \sim -4}).
\end{equation}
Since the constant in (\ref{f NL 1}) has the order of
\begin{equation}
M_{\text{Pl}} \sim \mathcal{O} (10^{19}) \, \text{GeV}, \quad v \sim M \sim \mathcal{O} (10^{15}) \, 
\text{GeV}, \quad \tilde{p}_{\zeta}\sim 6.1 \times 10^{-9}
\end{equation}
respectively, the non-Gaussianity is found as 
\begin{equation}
f_{\text{NL}} \sim \mathcal{O}(10^{-4\sim-1}) \times \alpha, \label{NG pipipi1}
\end{equation}
where $\alpha$ represents the similar contributions from other components such as $\Sigma_{22}, \Sigma_{33}$ and $\dot{\Sigma}_{11}$
and the additional group theoretical numerical factor that appears in extending to larger GUT gauge groups.
The order is $\alpha \sim \mathcal{O}(10^{0\sim1})$.Since the current observation limit (Fig.\ref{Fig4}) is
\begin{equation}
|f_{\text{NL}}|\lesssim 1,
\end{equation}
this result is consistent with the observation 
and it might be possible to detect the signature of the Higgs boson in GUT 
 by 21cm spectrum, future LSS and future CMB depending on our model parameters. 

Next, we consider interactions involving time derivatives. The sum of the graphs of $(+, -)$ and $(-, +)$ is given as follows:
\begin{eqnarray}
\langle \pi \pi \pi \rangle &=& \int_{-(1+ i\epsilon) \infty}^{0} a(\tau_{1})d\tau_{1}\,\int_{-(1+ i\epsilon) \infty}^{0} 
a(\tau_{2})d\tau_{2} (-i) a^{3}(\tau_{1}) A i a^{3}(\tau_{2}) A \nonumber \\
&& \times G_{+-}(\tau_{1}, \tau_{2}, p_{3}) \frac{1}{a(\tau_{1})} \dot{\Delta}_{+-}(0, \tau_{1}, p_{1}) 
\frac{1}{a(\tau_{1})} \dot{\Delta}_{+-}(0, \tau_{1}, p_{2}) \frac{1}{a(\tau_{2})} \dot{\Delta}_{++}(\tau_{2}, 0, p_{3}) \nonumber \\
&& + (p_{1}\leftrightarrow p_{3}) +  (p_{2}\leftrightarrow p_{3}) + \text{c.c.} 
\label{dpi3}
\end{eqnarray}
where $A$ is defined as
\begin{equation}
A:= 10 M^{2} v  - \left( 600\lambda_{1} + 140\lambda_{2} \right) v^{3}.
\end{equation}
Differentiating the propagation function $\Delta_{++}\left(\tau_{2}, 0, p\right)$ with respect to the first time yields
\begin{eqnarray}
\dot{\Delta}_{++}\left(\tau_{2}, 0, p\right) 
&=&\dot{\Delta}_{>}\left(\tau_{2}, 0, p\right) \theta\left(\tau_{2}-0\right) 
+ \Delta_{>}\left(\tau_{2}, 0, p\right) \delta\left(\tau_{2}-0\right)
 \nonumber \\
 && +\dot{\Delta}_{<}\left(\tau_{2}, 0, p\right) \theta\left(0-\tau_{2}\right) 
 - \Delta_{<}\left(\tau_{2}, 0, p\right) \delta\left(0-\tau_{2}\right) \nonumber \\
 &=&  \frac{ 1+ ic_{s}p \tau_{2}}{4 \epsilon M_{\text{Pl}} c_{s} p^{3} } 
 e^{- i c_{s} p  \tau_{2}} \delta(\tau_{2}) \nonumber \\
 && + \left(   \frac{- ic_{s}p }{4 \epsilon M_{\text{Pl}} c_{s} p^{3} } 
 e^{ i c_{s} p  \tau_{2}} + ic_{s}p \frac{ 1- ic_{s}p \tau_{2}}{4 \epsilon M_{\text{Pl}} c_{s} p^{3} } e^{ i c_{s} p  \tau_{2}}  \right) 
 \theta(-\tau_{2}) - \frac{ 1- ic_{s}p \tau_{2}}{4 \epsilon M_{\text{Pl}} c_{s} p^{3} } e^{ i c_{s} p  \tau_{2}} \delta(-\tau_{2}) 
 \nonumber \\
 &=& \frac{ 1+ ic_{s}p \tau_{2}}{4 \epsilon M_{\text{Pl}} c_{s} p^{3} } 
 e^{- i c_{s} p  \tau_{2}} \delta(\tau_{2}) - \frac{ 1- ic_{s}p \tau_{2}}{4 \epsilon M_{\text{Pl}} c_{s} p^{3} } 
 e^{ i c_{s} p  \tau_{2}} \delta(-\tau_{2}) +    \frac{c_{s}\tau_{2} }{4 \epsilon M_{\text{Pl}}  p } 
 e^{ i c_{s} p  \tau_{2}}  \theta(-\tau_{2}),
\end{eqnarray}
which leads to a simple form
\begin{equation}
\dot{\Delta}_{++}\left(\tau_{2}, 0, p\right) =  \frac{c_{s}\tau_{2} }{4 \epsilon M_{\text{Pl}}  p } e^{ i c_{s} p  \tau_{2}} 
\end{equation}
because of the $\tau_{2}$-integral. Similarly, differentiating the propagation function 
$\Delta_{+-}(0, \tau_{1}, p)$ with respect to the second time yields
\begin{eqnarray}
\dot{\Delta}_{+-}(0, \tau_{1}, p) = \dot{\Delta}_{<}(0, \tau_{1}, p) &=& \frac{ ic_{s}p
  }   { 4 \epsilon M_{\text{Pl}} c_{s} p^{3} } e^{- i c_{s} p  \tau_{1}} + (-ic_{s}p) \frac{ 1+ ic_{s}p \tau_{1} 
 }   { 4 \epsilon M_{\text{Pl}} c_{s} p^{3} } e^{- i c_{s} p  \tau_{1}} \nonumber \\
 &=& \frac{ c_{s}\tau_{1}}{4 \epsilon M_{\text{Pl}}  p } e^{- i c_{s} p  \tau_{1}}.
\end{eqnarray}
Substituting these results, $\langle \pi \pi \pi \rangle$ becomes
\begin{eqnarray}
\langle \pi \pi \pi \rangle &=& \int_{-(1+ i\epsilon) \infty}^{0} a(\tau_{1})d\tau_{1}\,\int_{-(1+ i\epsilon) \infty}^{0} 
a(\tau_{2})d\tau_{2} (-i) a^{3}(\tau_{1})A i a^{3}(\tau_{2}) A G_{+-}(\tau_{1}, \tau_{2}, p_{3}) \nonumber \\
&& \frac{1}{a(\tau_{1})} \dot{\Delta}_{+-}(0, \tau_{1}, p_{1}) \frac{1}{a(\tau_{1})} 
\dot{\Delta}_{+-}(0, \tau_{1}, p_{2}) \frac{1}{a(\tau_{2})} \dot{\Delta}_{++}(\tau_{2}, 0, p_{3}) + (p_{1}\leftrightarrow p_{3}) 
+  (p_{2}\leftrightarrow p_{3}) + \text{c.c.} \nonumber \\
&=& \int d\tau_{1}\, \int d\tau_{2}\, \left( - \frac{1}{H\tau_{1}}  \right)^{2} \left( - \frac{1}{H\tau_{2}}  \right)^{3} A^{2}  
\left(-\frac{ \pi }{ 4}\right)  e^{-\pi \text{Im}(\nu)}  H^{2} \left(  \tau_{1} \tau_{2} \right) ^{3/2} 
H^{(1)}_{\nu}(-p\tau_{2}) H^{(2)}_{\nu}(-p\tau_{1}) \nonumber \\
&&\times  \frac{c_{s}\tau_{1} }{4 \epsilon M_{\text{Pl}}  p_{1} } e^{- i c_{s} p_{1}  \tau_{1}}  
\frac{c_{s}\tau_{1} }{4 \epsilon M_{\text{Pl}}  p_{2} } e^{ -i c_{s} p_{2}  \tau_{1}}  
\frac{c_{s}\tau_{2} }{4 \epsilon M_{\text{Pl}}  p_{3} } e^{ i c_{s} p_{3}  \tau_{2}}  
\nonumber \\
&&+ (p_{1}\leftrightarrow p_{3}) +  (p_{2}\leftrightarrow p_{3}) + \text{c.c.}
\end{eqnarray}
Furthermore, using horizon exit approximations, we obtain
\begin{eqnarray}
\langle \pi \pi \pi \rangle
&\sim& \int^{\tau_{*}}_{-(1+i\epsilon)\infty} d\tau_{1}\, \int^{\tau_{*}}_{-(1+i\epsilon)\infty} 
d\tau_{2}\, \frac{-1}{H^{5} \tau^{2}_{1} \tau^{3}_{2}} A ^{2} \nonumber \\
&&  \times \left(-\frac{ \pi }{ 4}\right)   H^{2} \left(  \tau_{1} \tau_{2} \right) ^{3/2} (-i) 
\frac{2^{\nu}}{\pi} (-p_{3}\tau_{2})^{-\nu} \Gamma(\nu)  i \frac{2^{\nu}}{\pi} 
(-p_{3}\tau_{1})^{-\nu} \Gamma(\nu) \left(\frac{c_{s} }{4 \epsilon M_{\text{Pl}} } \right)^{3}  
\frac{1}{p_{1}p_{2}p_{3}} \tau^{2}_{1} \tau_{2}  \nonumber \\
&&+ (p_{1}\leftrightarrow p_{3}) +  (p_{2}\leftrightarrow p_{3}) + \text{c.c.} 
\nonumber \\
&=& \frac{2^{2\nu}}{4\pi} \Gamma^{2}(\nu) \frac{1}{H^{3}} \left(\frac{c_{s} }{4 \epsilon M_{\text{Pl}} } \right)^{3}  
A ^{2} \frac{1}{p_{1}p_{2}p_{3}}  p^{-2\nu}_{3} \nonumber \\
&& \times \int^{\tau_{*}}_{-(1+i\epsilon)\infty} d\tau_{1}\, \int^{\tau_{*}}_{-(1+i\epsilon)\infty} d\tau_{2}\, \tau^{3/2}_{1} \tau^{-1/2}_{2} (\tau_{1}\tau_{2})^{-\nu}.
\end{eqnarray}
Here, if we introduce the UV cutoff
\begin{equation}
\tau_{\Lambda} = -\Lambda^{-1},
\end{equation}
we can calculate as
\begin{eqnarray}
\langle \pi \pi \pi \rangle
&\sim& \frac{2^{2\nu}}{4\pi} \Gamma^{2}(\nu) \frac{1}{H^{3}} \left(\frac{c_{s} }{4 \epsilon M_{\text{Pl}} } \right)^{3}  
A ^{2} \frac{1}{p_{1}p_{2}p_{3}}  p^{-2\nu}_{3}  \left.\frac{1}{5/2- \nu} \tau_{1}^{5/2 - \nu} \right|^{\tau_{*}}_{\tau_{\Lambda}} \left. 
\frac{1}{1/2- \nu} \tau_{2}^{1/2 - \nu} \right|^{\tau_{*}}_{\tau_{\Lambda}} \nonumber \\
&&
\end{eqnarray}
Since the cutoff scale has a dimension of momentum, we can write
\begin{equation}
\Lambda = \gamma p, 
\end{equation}
using the dimensionless parameter $\gamma$. Therefore we can obtain
\begin{equation}
 \left.\frac{1}{5/2- \nu} \tau_{1}^{5/2 - \nu} \right|^{\tau_{*}}_{\tau_{\Lambda}} \left. \frac{1}{1/2- \nu} \tau_{2}^{1/2 - \nu} \right|^{\tau_{*}}_{\tau_{\Lambda}} =  \frac{4}{\left(5- 2\nu \right)  \left(1- 2\nu \right)  } (-p)^{2\nu-3} \left( 1- \gamma^{\nu-1/2}  - \gamma^{\nu-5/2} + \gamma^{2\nu-3}    \right).
\end{equation}
From the definition, non-Gaussianity is computed as
\begin{eqnarray}
f_{\text{NL}} 
&=& \frac{10}{9} \frac{1}{(2\pi)^{7}\tilde{p}^{2}_{\zeta}} p^{6} M^{-3}_{\text{Pl}}  (-H)^{3}\langle \pi \pi \pi \rangle \nonumber \\
&=&  \frac{10}{9} \frac{1}{(2\pi)^{7}\tilde{p}^{2}_{\zeta}} p^{6} M^{-3}_{\text{Pl}}  (-H)^{3} \nonumber \\
&& \times     \frac{2^{2\nu}}{4\pi} \Gamma^{2}(\nu) \frac{1}{H^{3}} \left(\frac{c_{s} }{4 \epsilon M_{\text{Pl}} } \right)^{3} A^{2} \frac{1}{p^{3}}  p^{-2\nu}  \nonumber \\
&& \times \frac{4}{\left(5- 2\nu \right)  \left(1- 2\nu \right)  } (-p)^{2\nu-3} 
\left( 1- \gamma^{\nu-1/2}  - \gamma^{\nu-5/2} + \gamma^{2\nu-3}    \right) \nonumber \\
&\sim& 10^{-12} \left(  \frac{c_{s}}{\epsilon}  \right)^{3} \left( 1- \gamma^{\nu-1/2}  - \gamma^{\nu-5/2} + \gamma^{2\nu-3}    \right) .
\end{eqnarray}
Now, recalling that
\begin{equation}
\epsilon c_{s} \sim 10^{-3\sim-4},
\end{equation}
we have
\begin{equation}
f_{\text{NL}} \sim 10^{-3\sim0} c^{6}_{s} \left( 1- \gamma^{\nu-1/2}  - \gamma^{\nu-5/2} + \gamma^{2\nu-3}    \right).
\end{equation}
From the viewpoint of effective field theory, the cutoff is the inflationary scale, and $\gamma$ is
\begin{equation}
 \gamma  \sim 1.
\end{equation}
In this case, non-Gaussianity has the width of
\begin{equation}
f_{\text{NL}} \sim 10^{-3\sim0} c^{6}_{s}.
\end{equation}
$c_{s}$ is a quantity such that it is 1 in the simplest model, and is not expected to vary significantly in order estimation. This result is comparable to that of case (\ref{Hankel approximation2}) where the time derivative is not included. The reasons are shown in the following table. 
\begin{center}
\begin{table}[!h]
        \centering
        
\begin{tabular}{|c|c|c|}
\hline 
  & factor & propagator \\
\hline 
 $\langle \pi \pi \pi \rangle$ & $\frac{2}{9}AH^{2}\left(  1+ \frac{\epsilon}{3} \right) \times 3AH\left(  1+ \frac{\epsilon}{3} \right) $ & $G_{+-} \Delta_{+-} \Delta_{+-} \Delta_{++}$ \\
\hline 
 $\langle \pi \pi \dot{\pi} \rangle$ & $\frac{2}{9}AH^{2}\left(  1+ \frac{\epsilon}{3} \right) \times A $ & $G_{+-} \Delta_{+-} \Delta_{+-} \dot{\Delta}_{++}$ \\
 & $AH\left(  3+ 2\epsilon \right) \times 3AH\left(  1+ \frac{\epsilon}{3} \right) $ & $G_{+-} \Delta_{+-} \dot{\Delta}_{+-} \Delta_{++}$
\\\hline 
 $\langle \pi \dot{\pi} \dot{\pi} \rangle$ & $A \times 3AH\left(  1+ \frac{\epsilon}{3} \right) $ & $G_{+-} \dot{\Delta}_{+-} \dot{\Delta}_{+-} \Delta_{++}$ \\
 & $AH\left(  3+ 2\epsilon \right) \times A $ & $G_{+-} \dot{\Delta}_{+-} \Delta_{+-} \dot{\Delta}_{++}$
\\\hline 
 $\langle \dot{\pi} \dot{\pi} \dot{\pi} \rangle$ & $A\times A$ & $G_{+-} \dot{\Delta}_{+-} \dot{\Delta}_{+-} \dot{\Delta}_{++}$ \\
 \hline
\end{tabular}
\caption{The factors and propagator for three point function of inflaton with various number of time derivatives.} 
        \end{table}
\end{center}
Table 1 shows the coefficients and propagator for three point function of inflaton 
with various number of time derivatives. 
In the table, the case where the three point function of inflaton without 
a time derivative is denoted by $\langle \pi \pi \pi \rangle$, 
with one, two and three time derivatives by $\langle \pi \pi \dot{\pi} \rangle$ , 
$\langle \pi \dot{\pi} \dot{\pi} \rangle$, and 
$\langle \dot{\pi} \dot{\pi} \dot{\pi} \rangle$. 
For example, $\langle \pi \pi \pi \rangle$ represents equation (\ref{pi3}) 
and $\langle \dot{\pi} \dot{\pi} \dot{\pi} \rangle$ represents equation (\ref{dpi3}). 
Taking into account that $a^{-1}=H\tau$ is added when performing the time derivative of the propagator, 
the coefficients are all of the form $A^{2}H^{3}$. 
The coefficients except for $A^{2}H^{3}$ of the interaction are all $\mathcal{O}(1)$, 
and the propagator only differs by about $\mathcal{O}(1)$ 
as long as the cutoff is set to $\Lambda = \gamma p$. 
Thus, they all give the same contribution to non-Gaussianity.

\begin{figure}[htbp]
 \begin{center}
  \includegraphics[width=5cm]{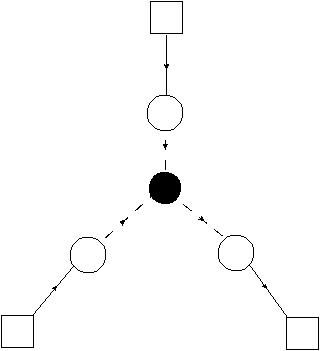}
 \end{center}
 \caption{The graph generated by the three-point interaction of $\Sigma$. 
 There are $2^{4}$ in total, and this is an example, representing the graph of $(-, -, -, +)$.}
 \label{Fig5}
 \end{figure}

Next, we consider the graph created by the three-point interaction of $\Sigma$ (Fig.\ref{Fig5}).

Let $\tau_{4}$ be the time for the part of 3-point interaction 
and $\tau_{1}, \tau_{2}, \tau_{3}$ be the time for the rest of $\pi\Sigma$ interaction. 
According to the rule of in-in formalism, there exist $2^{4}$ types of graphs. Consider the $(-, -, -, +)$ graphs 
and its complex conjugation $(+, +, +, -)$ in the order $(\tau_{1}, \tau_{2}, \tau_{3}, \tau_{4})$. 
The case where the interaction does not have time derivative is written down as
\begin{eqnarray}
\langle \pi \pi \pi \rangle &=& \int_{-(1+ i\epsilon) \infty}^{0} a(\tau_{1})d\tau_{1}\,\int_{-(1+ i\epsilon) \infty}^{0} 
a(\tau_{2})d\tau_{2}\int_{-(1+ i\epsilon) \infty}^{0} 
a(\tau_{3})d\tau_{3}\int_{-(1+ i\epsilon) \infty}^{0} 
a(\tau_{4})d\tau_{4} \nonumber \\
&& \times (-i) a(\tau_{1}) 3a^{3}(\tau_{1})AH \left( 1+ \frac{\epsilon}{3}  \right)\Delta_{+-}(0, \tau_{1}, p_{1})  
G_{-+}(\tau_{1}, \tau_{4}, p_{1}) \nonumber \\
&& \times (-i) a(\tau_{2})  3a^{3}(\tau_{2})AH \left( 1+ \frac{\epsilon}{3}  \right) \Delta_{+-}(0, \tau_{2}, p_{2}) 
G_{-+}(\tau_{2}, \tau_{4}, p_{2})\nonumber \\
&&\times  ia(\tau_{4}) (-1) a^{3}(\tau_{4}) 40\lambda _{1} v G_{+-}(\tau_{4}, \tau_{3}, p_{3}) \nonumber \\
&& \times (-i) a(\tau_{3})  3a^{3}(\tau_{3})AH \left( 1+ \frac{\epsilon}{3} \right) \Delta_{-+}(\tau_{3}, 0, p_{3})
  + (p_{1}\leftrightarrow p_{3}) +  (p_{2}\leftrightarrow p_{3}) + \text{c.c.} \nonumber \\
\end{eqnarray}
Substituting the propagator, we obtain
\begin{eqnarray}
\langle \pi \pi \pi \rangle &=& 40\times 3^{3}\lambda _{1} v \left( 1+ \frac{\epsilon}{3} \right)^{3}A^{3}H^{-7} \left(  \frac{ 1
  }   { 4 \epsilon M_{\text{Pl}} c_{s}  }   \right)^{3}  \left(-\frac{ \pi }{ 4}\right)^{3}  
  e^{-3\pi \text{Im}(\nu)} \left(  \frac{1}{p_{1}p_{2}p_{3}} \right)^{3} \nonumber \\
&&\times \int_{-(1+ i\epsilon) \infty}^{0} d\tau_{1}\,\int_{-(1+ i\epsilon) \infty}^{0} 
d\tau_{2}\int_{-(1+ i\epsilon) \infty}^{0} 
d\tau_{3}\int_{-(1+ i\epsilon) \infty}^{0} 
d\tau_{4}  \left(1+ ic_{s}p_{1} \tau_{1}\right) \left(1+ ic_{s}p_{2} \tau_{2}\right) 
\left(1+ ic_{s}p_{3} \tau_{3}\right) \nonumber \\
&&\times e^{- i c_{s} p_{1}  \tau_{1}}  e^{- i c_{s} p_{2}  \tau_{2}}   e^{- i c_{s} p_{3}  \tau_{3}}  
\left( \tau_{1}\tau_{2}\tau_{3}  \right)^{-5/2} \tau_{4}^{1/2} \nonumber \\
&&\times H^{(1)}_{\nu}(-p_{1}\tau_{1}) H^{(1)}_{\nu}(-p_{2}\tau_{2})H^{(1)}_{\nu}(-p_{3}\tau_{3}) 
H^{(2)}_{\nu}(-p_{1}\tau_{4})H^{(2)}_{\nu}(-p_{2}\tau_{4}) H^{(2)}_{\nu}(-p_{3}\tau_{4}) \nonumber \\
&& + (p_{1}\leftrightarrow p_{3}) +  (p_{2}\leftrightarrow p_{3}) + \text{c.c.}
\end{eqnarray}
Using horizon exit approximations, we can compute
\begin{eqnarray}
\langle \pi \pi \pi \rangle &\simeq& - 6\times 40\times 3^{3}\lambda _{1} v 
\left( 1+ \frac{\epsilon}{3} \right)^{3}A^{3}H^{-7} 
\left(  \frac{ 1
  }   { 4 \epsilon M_{\text{Pl}} c_{s}  }   \right)^{3}  \left(\frac{ \pi }{ 4}\right)^{3} \left(\frac{ 2^{\nu} }{ \pi}\right)^{6} 
  \Gamma^{6}(\nu) \left(  \frac{1}{p_{1}p_{2}p_{3}} \right)^{3} \nonumber \\
&&\times \int_{-(1+ i\epsilon) \infty}^{\tau_{1*}}d\tau_{1}\,\int_{-(1+ i\epsilon) \infty}^{\tau_{2*}} 
d\tau_{2}\int_{-(1+ i\epsilon) \infty}^{\tau_{3*}} 
d\tau_{3}\int_{-(1+ i\epsilon) \infty}^{\tau_{4*}} 
d\tau_{4}  \nonumber \\
&& \times \left( \tau_{1}\tau_{2}\tau_{3}  \right)^{-5/2} \tau_{4}^{1/2} \left( -p_{1}\tau_{1}\right)^{-\nu}\left( -p_{2}\tau_{2}\right)^{-\nu}\left( -p_{3}\tau_{3}\right)^{-\nu}\left( -p_{1}\tau_{4}\right)^{-\nu}\left( -p_{2}\tau_{4}\right)^{-\nu}\left( -p_{3}\tau_{4}\right)^{-\nu} \nonumber \\
\end{eqnarray}
The first factor 6 comes from momentum exchange and contributions from complex conjugation. 
Now, if we insert the cutoff $\Lambda = \gamma p$ for $\tau_{4}$, we obtain
\begin{eqnarray}
\langle \pi \pi \pi \rangle &\simeq&  6\times 40\times 3^{3}\lambda _{1} v 
\left( 1+ \frac{\epsilon}{3} \right)^{3}A^{3}H^{-7} \left(  \frac{ 1
  }   { 4 \epsilon M_{\text{Pl}} c_{s}  }   \right)^{3}  \left(\frac{ \pi }{ 4}\right)^{3} 
  \left(\frac{ 2^{\nu} }{ \pi}\right)^{6} \Gamma^{6}(\nu) \left(  \frac{1}{p_{1}p_{2}p_{3}} \right)^{3} \nonumber \\
&&\times \left( \frac{2} {3 +  2\nu}  \right)^{3} \frac{2}{3-6\nu}  (p_{1}p_{2})^{3/2 - \nu}  
(-1)^{3\nu -3/2} p_{3}^{2\nu}\left( 1- \gamma^{3\nu-3/2}    \right),
\end{eqnarray}
thus non-Gaussianity is given by
\begin{eqnarray}
f_{\text{NL}} 
&=& \frac{10}{9} \frac{1}{(2\pi)^{7}\tilde{p}^{2}_{\zeta}} p^{6} M^{-3}_{\text{Pl}}  
(-H)^{3}\langle \pi \pi \pi \rangle  \nonumber \\
&=&  \frac{10}{9} \frac{1}{(2\pi)^{7}\tilde{p}^{2}_{\zeta}} p^{6} M^{-3}_{\text{Pl}}  
(-H)^{3} \nonumber \\
&& \times    6\times 40\times 3^{3}\lambda _{1} v A^{3}H^{-7} 
\left(  \frac{ 1
  }   { 4 \epsilon M_{\text{Pl}} c_{s}  }   \right)^{3}  \left(\frac{ \pi }{ 4}\right)^{3} 
  \left(\frac{ 2^{\nu} }{ \pi}\right)^{6} \Gamma^{6}(\nu) 
  \left(  \frac{1}{p_{1}p_{2}p_{3}} \right)^{3} \nonumber \\
&&\times \left( \frac{2} {3 +  2\nu}  \right)^{3} \frac{2}{3-6\nu}  
(p_{1}p_{2})^{3/2 - \nu}  (-1)^{3\nu -3/2} p_{3}^{2\nu}\left( 1- \gamma^{3\nu-3/2}    \right).
\end{eqnarray}
Substituting numerical values as before gives the result
\begin{equation}
| f_{\text{NL}}  | = \mathcal{O}(10^{-3\sim0}).
\end{equation}
The calculations for the case where the interaction involves time derivatives can be performed similarly to the discussion in Table 1, 
and they all yield the same contribution to non-Gaussianity.

From the above, non-Gaussianity has the form
\begin{equation}
| f_{\text{NL}}  | = \mathcal{O}(10^{-3\sim0}) \times \alpha
\end{equation}
in summary. $\alpha$ is the contribution from components other than $\Sigma_{11}$, 
or the additional group theoretical numerical factor that appears in extending to larger GUT gauge groups, 
and so on, that is the quantity summarizes the factors that bring about the same level of contribution, 
and has an order of magnitude of $\mathcal{O}(10^{0\sim1})$.

\section{Conclusion}
The Standard Model of elementary particles, which has successfully explained many physical phenomena, 
 is probably one of the most successful physical theories. 
However, the nature is full of rich phenomena that cannot be explained by the Standard Model alone.  
GUT is one of the attempts to describe these interesting phenomena. 
GUT is a fascinating theory that unifies the three interactions that exist in nature: 
strong interaction, electromagnetic interaction, and weak interaction. 
Although many researchers have tried to verify it, it has not yet been confirmed. 
For example, we have been looking for proton decay in Super Kamiokande, but have not obtained that reaction. 
In addition, it is difficult to investigate by using accelerators because the GUT scale is very high energy ($10^{15}$GeV). 
For this reason, Cosmological Collider Physics has been the focus of much attention in recent years. 
Cosmological Collider Physics is a method to obtain information on elementary particles by using the effective field theory of inflation. 
Quantum fluctuations generated in the short time after the birth of the universe are stretched by inflation. 
It appears in the form of non-Gaussianity by observing the cosmic microwave background radiation. 
This means that Cosmological Collider Physics is a very interesting way to obtain information on high energy elementary particles that cannot be reached by terrestrial accelerators by means of precise observation of the universe. 

In this paper, we focus on the case where the energy scale of the inflation is close to the GUT scale, 
 and discuss if the GUT can be verified by calculating the non-Gaussianity due to the Higgs boson in GUT. 
Concretely, in addition to the effective action of inflation, we considered the action of the adjoint Higgs scalar field in SU(5) GUT. 
A characteristic feature of this model is that the Higgs boson has a vacuum expectation value 
 due to spontaneous symmetry breaking, which leads to linear interactions of Higgs boson with the inflation. 
Using these interactions, the three point function of the inflatons is generated by the tree level exchange of the adjoint Higgs boson. 
The graphs contributing to the inflaton three point function can be computed by performing horizon exit approximation, 
 and non-Gaussianity is evaluated from the obtained values. 
As a result, we have shown 
\begin{equation}
|f_{\text{NL}}| \lesssim 1
\end{equation}
for non-Gaussianity without a drastic fine-tuning of parameters. 
This result is consistent with the current observed limit and suggests the existence of the adjoint Higgs boson in GUT 
and it might be possible to detect the signature of the Higgs boson in GUT 
 by 21cm spectrum, future LSS and future CMB depending on our model parameters. 
 
\appendix
\section{Calculation of $\sqrt{-g}$}
For the determinant of the metric, 
 starting from the curvature fluctuation in unitary gauge $\zeta(t, \bm{x})$, 
 we derive inflaton 
We can write determinant of the metric in unitary gauge as
\begin{equation}
\sqrt{-g} = Na^{3}(t) = \left(  1+ \frac{1}{H(t)} \frac{d}{dt} \zeta(t) \right) a^{3}(t) 
\label{metric in unitary gauge}
\end{equation}
using ADM formalism. 
Curvature fluctuations $\zeta(t, \bm{x})$ and NG boson $\pi(\tilde{t}, \bm{x})$ are related by the relation
\begin{equation}
\zeta(t, \bm{x}) = -H(\tilde{t}) \pi(\tilde{t}, \bm{x})
\end{equation}
by the transformation of time
\begin{equation}
t \mapsto \tilde{t} = t - \pi(\tilde{t}, \bm{x}).
\end{equation}
By rewriting the derivative of $\zeta(t, \bm{x})$ with respect to $t$ into the derivative with respect to $\tilde{t}$, 
we obtain
\begin{eqnarray}
\frac{d}{dt} \zeta(t, \bm{x})  &=& \frac{d \tilde{t}}{d t} \frac{d}{d \tilde{t}} \left(-H(\tilde{t}) \pi(\tilde{t}, \bm{x}) \right) \nonumber \\
&=& - \left[ 1- \frac{d \tilde{t}}{d t}  \dot{\pi}(\tilde{t}) \right] \left(\dot{H} \pi + H \dot{\pi}  \right)  \nonumber \\
&=& - \left( 1-\dot{\pi} + \dot{\pi}^{2}  \right) \left(\dot{H} \pi + H \dot{\pi}  \right) \nonumber \\
&=& \dot{H}(\tilde{t}) \left( -\pi + \pi \dot{\pi} \right) + H(\tilde{t}) \left(- \dot{\pi} + \dot{\pi}^{2}  \right) 
\end{eqnarray}
up to the second order of $\pi$, 
 where the dot represents the derivative with respect to $\tilde{t}$. 
After applying the time transformation to $H^{-1}(t)$, we obtain
\begin{equation}
H^{-1}(t) = H^{-1}(\tilde{t} + \pi) = \left(  H(\tilde{t}) + \dot{H}(\tilde{t}) \pi + \frac{1}{2} \ddot{H}(\tilde{t}) \pi^{2}  \right)^{-1}
\end{equation}
which can be written as
\begin{eqnarray}
H^{-1}(t) &\simeq& \left(  H(\tilde{t}) + \dot{H}(\tilde{t}) \pi  \right)^{-1} \nonumber \\
&=& \frac{1}{H(\tilde{t})} \left( 1+ \frac{\dot{H}(\tilde{t})}{H(\tilde{t})} \pi\right)^{-1} \nonumber \\
&=&  \frac{1}{H(\tilde{t})} \left( 1- H(\tilde{t}) \epsilon(\tilde{t}) \pi\right)^{-1} \nonumber \\
&=& \frac{1}{H(\tilde{t})} + \epsilon(\tilde{t}) \pi + H\epsilon^{2} \pi^{2}
\end{eqnarray}
Note that $\ddot{H}$ is in the second order of $\epsilon= -\dot{H}/H^{2}$. 
Similarly, applying the time transformation to the scale factor $a(t)$, 
 we obtain
\begin{eqnarray}
a^{3}(t) = a^{3}( \tilde{t} + \pi)  &=& \left[  a(\tilde{t}) + \dot{a}(\tilde{t}) \pi + \frac{1}{2} \ddot{a}(\tilde{t}) \pi^{2} \right]^{3} \nonumber \\
&=& a^{3}(\tilde{t}) + 3 a^{2}(\tilde{t}) \dot{a}(\tilde{t}) \pi (\tilde{t}) + \frac{3}{2} a^{2} \ddot{a} \pi^{2} + 3a \dot{a}^{2} \pi^{2}.
\end{eqnarray}
Putting them together, the determinant of the metric (\ref{metric in unitary gauge}) expands to
\begin{eqnarray}
\sqrt{-g} &=& \left(  1+ \frac{1}{H(t)} \frac{d}{dt} \zeta(t, \bm{x}) \right) a^{3}(t) \nonumber \\
&=& \left[ 1+ \frac{\dot{H}(\tilde{t})}{H(\tilde{t})} \left\{ -\pi + \pi \dot{\pi} \right\} - \dot{H} \epsilon \pi^{2} - \dot{\pi} + \dot{\pi}^{2} - H \epsilon \pi \dot{\pi} \right] \nonumber \\
&& \times \left[ a^{3}(\tilde{t}) + 3 a^{2}(\tilde{t}) \dot{a}(\tilde{t}) \pi (\tilde{t}) + \frac{3}{2} a^{2} \ddot{a} \pi^{2} + 3a \dot{a}^{2} \pi^{2}  \right] \nonumber \\
&=& a^{3}(\tilde{t}) + 3a^{2} \dot{a} \pi + \frac{3}{2} a^{2} \ddot{a} \pi^{2} + 3a \dot{a}^{2} \pi^{2} - \frac{\dot{H}}{H} \left( a^{3} \pi  + 3a^{2} \dot{a} \pi^{2}\right) + \frac{\dot{H}}{H} a^{3} \pi \dot{\pi} \nonumber \\
&{}&  - \dot{H} \epsilon a^{3} \pi^{2}  - a^{3} \dot{\pi}  - 3a^{2} \dot{a} \pi \dot{\pi}  + a^{3} \dot{\pi}^{2}  - H \epsilon a^{3} \pi \dot{\pi}.
\end{eqnarray}
Using the relation
\begin{equation}
\dot{a} = aH , \quad \ddot{a} = aH^{2} + a \dot{H},
\end{equation}
and expressing the derivative of the scale factor $a$ in terms of the Hubble $H$, 
we can obtain
\begin{eqnarray}
\sqrt{-g} &=& a^{3} (\tilde{t}) + 3a^{3}H\pi + \frac{3}{2} a^{3}\left( H^{2} + \dot{H} \right)\pi^{2} + 3a^{3} H^{2} \pi^{2}\nonumber \\
&{}&  -\frac{ \dot{H}} {H} a^{3} \pi - 3a^{3} \dot{H} \pi^{2} + \frac{ \dot{H}} {H} a^{3} \pi \dot{\pi} 
- \dot{H} \epsilon a^{3} \pi^{2} - a^{3} \dot{\pi} - 3a^{3} H \pi \dot{\pi} + a^{3} \dot{\pi}^{2} - H \epsilon a^{3} \pi \dot{\pi} \nonumber \\
 &=& a^{3}(\tilde{t}) + 3a^{3} H \left(  1+ \frac{\epsilon}{3}\right) \pi - a^{3} \dot{\pi} 
 + a^{3} H^{2} \left( \frac{9}{2} + \frac{3}{2} \epsilon \right) \pi^{2} - a^{3}H \left( 3+ 2\epsilon  \right) \pi \dot{\pi} 
 + a^{3} \dot{\pi}^{2}. \nonumber \\
 &{}&
\end{eqnarray}

\section{In-in formalism}

In this section, we briefly describe the propagators and Feynman diagram in the in-in formalism (Schwinger-Keldysh formalism), 
 which is a formalism describing non-equilibrium systems appearing in the inflationary universe. 
For details, please refer to \cite{SK}.

\begin{figure}[htbp]
 \begin{center}
  \includegraphics[width=100mm]{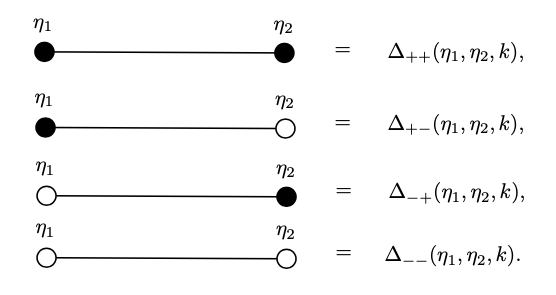}
 \end{center}
 \vspace*{-5mm}
 \caption{Graphical representation for an internal line in in-in formalism.}
\label{Fig6}
 \end{figure}

\begin{figure}[htbp]
 \begin{center}
  \includegraphics[width=100mm]{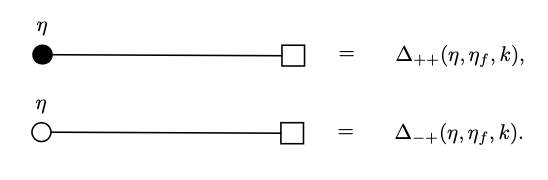}
 \end{center}
 \vspace*{-5mm}
 \caption{Graphical representation for an external line in in-in formalism.}
\label{Fig7}
 \end{figure}

For simplicity, we now take a real scalar field $\varphi$ as an example, 
but the generalization to other fields is straightforward.  
To consider the generating function for a real scalar field $\varphi$ in in-in formalism, 
 we need two fields $\varphi_{\pm}(\eta, \bm{x})$ and their corresponding sources $J_{\pm}(\eta, \bm{x})$.
\begin{equation}
Z\left[J_{+}, J_{-}\right]=\int \mathcal{D} \varphi_{+} \mathcal{D} \varphi_{-} 
 \exp \left[i \int_{\eta_{0}}^{\eta_{f}} \mathrm{d} \eta d^{3}x 
 \left(\mathscr{L}_{\mathrm{cl}}\left[\varphi_{+}\right]-\mathscr{L}_{\mathrm{cl}}\left[\varphi_{-}\right]
 +J_{+} \varphi_{+}-J_{-} \varphi_{-}\right)\right].
\end{equation}
This is the form of the generating functional in flat spacetime with the addition of the complex conjugate contribution. 
The reason for adding such contributions is that the expectation values of physical quantities in flat spacetime field theories 
 are calculated assuming that the initial vacuum and the final vacuum are in the same state, 
 whereas in inflationary spacetime the initial vacuum and the final vacuum are in different states. 
The physical meaning of this operation, which changes $i$ to $-i$, 
 is that the time evolution from the initial state to the final state and backward in time from there, 
 which implies that the correct time evolution contribution of the vacuum is taken into account. 
Once written in a form that incorporates this contribution, we can proceed in the same way as in flat spacetime field theories. 
Dividing  the Lagrangian into the free field part $\mathscr{L}_{0}$ and the interaction part $\mathscr{L}_{\mathrm{int}}$, 
 we express the generating functional as follows.
\begin{eqnarray}
Z\left[J_{+}, J_{-}\right] 
&=& \exp \left[i \int_{\eta_{0}}^{\eta_{f}} d \eta d^{3}x
 \left(\mathscr{L}_{\mathrm{int}}\left[\frac{\delta}{i \delta J_{+}}\right]
 -\mathscr{L}_{\mathrm{int}}\left[-\frac{\delta}{i \delta J_{-}}\right]\right)\right] Z_{0}\left[J_{+}, J_{-}\right], \\ 
Z_{0}\left[J_{+}, J_{-}\right] & \equiv &\int \mathcal{D} \varphi_{+} \mathcal{D} \varphi_{-} 
 \exp \left[i \int_{\eta_{0}}^{\eta_{f}} d \eta d^{3}x\left(\mathscr{L}_{0}\left[\varphi_{+}\right]
 -\mathscr{L}_{0}\left[\varphi_{-}\right]+J_{+} \varphi_{+}-J_{-} \varphi_{-}\right)\right]. \nonumber \\
&{}& 
\end{eqnarray}
The difference from the flat spacetime field theory is that there are four kinds of propagators, 
 according to the subscripts of $+$ and $-$.
\begin{equation}
-i G_{a b}\left(\eta_{1}, \bm{x}_{1} ; \eta_{2}, \bm{x}_{2}\right) \equiv 
 \left.\frac{\delta}{i a \delta J_{a}\left(\eta_{1}, \bm{x}_{1}\right)} 
 \frac{\delta}{i b \delta J_{b}\left(\eta_{2}, \bm{x}_{2}\right)} Z_{0}\left[J_{+}, J_{-}\right]\right|_{J_{\pm}=0},
\end{equation}
where $a$ and $b$ denote $a, b=\pm$. 
To give an example, a concrete propagator of type $(+, +)$ is
\begin{eqnarray}
-i G_{++}\left(\eta_{1}, \bm{x}_{1} ; \eta_{2}, \bm{x}_{2}\right) 
&=& \left.\frac{\delta}{i \delta J_{+}\left(\eta_{1}, \bm{x}_{1}\right)} 
 \frac{\delta}{i \delta J_{+}\left(\eta_{2}, \bm{x}_{2}\right)} 
 Z_{0}\left[J_{+}, J_{-}\right]\right|_{J_{\pm}=0} \nonumber \\
&=& \int \mathcal{D} \varphi_{+} \mathcal{D} \varphi_{-} 
 \varphi_{+}\left(\eta_{1}, \bm{x}_{1}\right) \varphi_{+}\left(\eta_{2}, \bm{x}_{2}\right) 
 e^{i \int d \eta d^{3}x\left(\mathscr{L}_{0}\left[\varphi_{+}\right]-\mathscr{L}_{0}\left[\varphi_{-}\right]\right)} 
 \nonumber \\
&=& \left\langle\Omega\left|\mathrm{T}\left\{\varphi\left(\eta_{1}, \bm{x}_{1}\right) 
 \varphi\left(\eta_{2}, \bm{x}_{2}\right)\right\}\right| \Omega\right\rangle,
\end{eqnarray}
where T is the time-ordered product operator. 
Using the translational and rotational symmetries at each time in the background spacetime, 
 we can perform the Fourier transformation and move to the three-dimensional momentum space. 
We can obtain the propagators in momentum space
\begin{equation}
\Delta_{a b}\left(\eta_{1}, \eta_{2}, k\right)
=-i \int d^{3}x e^{-i\bm{k} \cdot \bm{x}} G_{a b}\left(\eta_{1}, \bm{x} ; \eta_{2}, \bm{0}\right), 
\end{equation}
where $-i$ is added to simplify the Feynman rule in momentum space. 
Here $k=|\bm{k}|$ is the magnitude of the three-dimensional momentum, 
 and the propagator in momentum space 
 $\Delta_{a b}\left(\eta_{1}, \eta_{2}, k\right)$ is a function of magnitude due to the rotational symmetry, 
 independent of the direction of the three-dimensional momentum. 
The field $\varphi$ can be represented by a mode function $v(\eta, \bm{k})$ 
 and creation and annihilation operators for a given three-dimensional momentum $\bm{k}$. 
The four types of propagators are concretely written as
\begin{eqnarray}
\Delta_{++}\left(\eta_{1}, \eta_{2}, k\right) 
&=& \Delta_{>}\left(\eta_{1}, \eta_{2}, k\right) \theta\left(\eta_{1}-\eta_{2}\right)
 +\Delta_{<}\left(\eta_{1}, \eta_{2}, k\right) \theta\left(\eta_{2}-\eta_{1}\right),  \\
\Delta_{+-}\left(\eta_{1}, \eta_{2}, k\right)&=&\Delta_{<}\left(\eta_{1}, \eta_{2}, k\right),  \\
\Delta_{-+}\left(\eta_{1}, \eta_{2}, k\right)&=& \Delta_{>}\left(\eta_{1}, \eta_{2}, k\right),  \\
\Delta_{--}\left(\eta_{1}, \eta_{2}, k\right)
&=& \Delta_{<}\left(\eta_{1}, \eta_{2}, k\right) \theta\left(\eta_{1}-\eta_{2}\right)
 +\Delta_{>}\left(\eta_{1}, \eta_{2}, k\right) \theta\left(\eta_{2}-\eta_{1}\right),
\end{eqnarray}
where $\Delta_{>}\left(\eta_{1}, \eta_{2}, k\right)$ and $\Delta_{<}\left(\eta_{1}, \eta_{2}, k\right)$ 
 are defined by
\begin{eqnarray}
\label{delta>}
&&\Delta_{>}\left(\eta_{1}, \eta_{2}, k\right)  \equiv v\left(\eta_{1}, k\right) v^{*}\left(\eta_{2}, k\right) , \\ 
&&\Delta_{<}\left(\eta_{1}, \eta_{2}, k\right) \equiv v^{*}\left(\eta_{1}, k\right) v\left(\eta_{2}, k\right)
\label{delta<}
\end{eqnarray}
and $\theta(\eta)$ is a step function of $\eta$. 
To represent these four propagators in terms of a diagram, 
 we can use black and white circles for $+$ and $-$, respectively (Fig.\ref{Fig6}).
These are graphical rules for internal lines. 
The external line connects the time slice with end time $\eta = \eta_{f}$ (boundary point) to the other time slice. 
Since the boundary points do not have the distinction between $+$ and $-$, 
 there are only two types of propagators for the external lines, which are called as bulk-to-boundary propagators. 
To represent this boundary point, we use a square in a diagrams (Fig.\ref{Fig7}).

As an example, we consider the propagator of a complex scalar field $\phi$ with mass $m$ in the de Sitter spacetime. 
By using conformal time $\eta$ and the Minkowski metric $\eta^{\mu\rho}$ to write down the action, we have
\begin{equation}
S  =  \int d^{4}x\,
\left[ \eta^{\mu\rho}\phi^* \partial_{\mu}\partial_{\rho}\phi
- a^{2}m^{2}|\phi|^{2}
\right] .
\end{equation}
From this action, we obtain the equation of motion
\begin{equation}
\left( \eta^{\mu\rho}\partial_{\mu}\partial_{\rho} - a^{2}m^{2} \right)\phi =0 ,
\label{KKeq}
\end{equation}
which is the Klein-Gordon equation with mass $a^{2}m^{2}$. 
Due to the rotational symmetry, we can transform the scalar field $\phi$ into three dimensional momentum space
 and write its mode function as $u(\eta, k)$, 
 the Klein-Gordon equation (\ref{KKeq}) is
\begin{equation}
\frac{\partial^{2}u}{\partial \eta^{2}}(\eta, k) 
 + \left(k^{2} + \frac{m^{2}}{H^{2}\eta^{2}} \right)u(\eta, k) = 0 .
\end{equation}
The solution to this Klein-Gordon equation is
\begin{equation}
u(\eta, k) 
= -i\frac{\sqrt{\pi}}{2}e^{i\pi\left(\nu/2 +1/4  \right)} 
 \left( -\eta  \right)^{\frac{1}{2}} H_{\nu}(-k\eta) ,
\label{Hankel}
\end{equation}
where $ H_{\nu}(-k\eta)$ is a Hankel function of the first kind and we define the index
\begin{equation}
\nu \equiv \sqrt{\frac{1}{4} - \left(   \frac{m}{H} \right)^{2}  }.
\end{equation}
Therefore, a scalar field with mass $m^{2}$ in de Sitter spacetime has propagators
\begin{eqnarray}
G_{>}\left(\eta_{1}, \eta_{2}, k\right) 
&=& - i \frac{\sqrt{\pi}}{2} e^{i\pi \left( \nu/2 + 1/4 \right)} H \left(  -\eta_{1} \right) ^{3/2} H^{(1)}_{\nu}(-k\eta_{1}) \times 
\nonumber \\
&&  i \frac{\sqrt{\pi}}{2} e^{i\pi \left( \nu^{*}/2 + 1/4 \right)} H \left(  -\eta_{2} \right) ^{3/2} H^{(2)}_{\nu}(-k\eta_{2}) \nonumber \\
&=& -\frac{ \pi }{ 4} e^{-\pi \text{Im}(\nu)}  H^{2} \left(  \eta_{1} \eta_{2} \right) ^{3/2} H^{(1)}_{\nu}(-k\eta_{1}) H^{(2)}_{\nu}(-k\eta_{2}),  \\
G_{<}\left(\eta_{1}, \eta_{2}, k\right) 
&=& -\frac{ \pi }{ 4} e^{-\pi \text{Im}(\nu)}  H^{2} \left(  \eta_{1} \eta_{2} \right) ^{3/2} H^{(1)}_{\nu}(-k\eta_{2}) H^{(2)}_{\nu}(-k\eta_{1})
\end{eqnarray}
from eq.(\ref{delta>}) and eq.(\ref{delta<}).


\end{document}